
\def\square{\kern1pt\vbox{\hrule height 1.2pt\hbox{\vrule width 1.2pt\hskip 3pt
   \vbox{\vskip 6pt}\hskip 3pt\vrule width 0.6pt}\hrule height 0.6pt}\kern1pt}
\magnification 1200

\voffset=-.1in
\vsize=7.5in
\hsize=5.6in
\tolerance 10000

\baselineskip 12pt plus 1pt minus 1pt
\pageno=0
\centerline{\bf REPLY TO}
\medskip
\centerline{\bf ``COMMENT ON `GRAVITY AND THE POINCAR\'E GROUP'
'' }
\smallskip
\smallskip
\vskip 24pt
\centerline{ G. Grignani}
\vskip 8pt
\centerline{\it Dipartimento di Fisica}
\centerline{\it Universit\'a degli Studi di Perugia}
\centerline{\it I-06100  Perugia -- ITALY}
\vskip 12pt
\centerline{and}
\vskip 12pt
\centerline{ G. Nardelli}
\vskip 8pt
\centerline{\it Dipartimento di Fisica}
\centerline{\it Universit\'a degli Studi di Trento}
\centerline{\it I-38050  Povo (TN) -- ITALY}
\vskip 48pt
\centerline{\bf Abstract}
\bigskip
In the first order form, the model considered by Strobl presents, besides
local Lorentz and diffeomorphism invariances, an additional local non-linear
symmetry. When the model is realized as a Poincar\'e gauge theory according
to the procedure outlined in Refs.[1,2], the generators of the non-linear
symmetry are responsible for the ``nasty
constraint algebra''. We show that not only the Poincar\'e gauge theoretic
formulation of the model is  not the cause  of the emerging of the undesirable
constraint algebra, but  actually allows to overcome the problem.  In fact one
can fix the additional symmetry without breaking the Poincar\'e gauge
symmetry and the diffeomorphisms,
so that, after a preliminary Dirac procedure, the remaining
constraints uniquely satisfy the Poincar\'e algebra. After the additional
symmetry is fixed, the equations of motion are unaltered.  The objections to
our method raised by Strobl in Ref.[3] are then
immaterial. Some minor
points put forward in Ref.[3] are also discussed.

\vfill
\vskip -12pt
\noindent  DFUPG-72-1993

\noindent UTF-290-1993

\hfill March 1993

\eject
\baselineskip 12pt plus 1pt minus 1pt

\medskip
\nobreak
In Refs.[1,2] we showed that gravity in any dimension
with all its couplings to matter  can be formulated as a gauge theory
of the Poincar\'e group. In order to reach this goal, a set of auxiliary
fields - the Poincar\'e coordinates $q^a (x)$ - are introduced, and the
Poincar\'e covariant derivative of such fields in the defining
representation is identified with the {\it zweibein}
$V^a{}_\mu$,\footnote{$^\dagger$}{For later convenience, we shall
consider here the 2 dimensional case.
For the 3 and 4 dimensional case see Ref. [1].
The conventions on the indices and
on the antisymmetric symbols are those of Ref.  [2].}
$$ {\cal D}_\mu q^a (x) = \partial_\mu q^a + \omega_\mu
\varepsilon^a{}_b q^b
+e^a{}_\mu \equiv V^a{}_\mu\ \ , \eqno(1)$$
so that the metric turns out Poincar\'e gauge invariant.

The spin connection $\omega_\mu$ and $e^a{}_\mu$ can be combined
together as components of the $ISO(1,1)$ gauge potential,
$A_\mu =  J \omega_\mu + P_a e^a{}_\mu$, where
$J$ and $P_a$ are the Lorentz and the translation generators,
respectively.

Recently  Strobl wrote a Comment [3]  criticizing our
papers [1,2].
 Besides some minor points that we shall
briefly discuss below, the main  criticism made in Ref.[3] is that if our
method to reformulate gravitational theories as Poincar\'e gauge theories is
applied to a given 2 dimensional model of non-Einsteinian gravity, one finds a
``nasty''
constraint algebra (with structure functions instead of constants)
that is not a representation of the gauge group.

In this reply we shall show that the above  conclusion is mistaken
and that one can suitably apply a canonical  Dirac procedure so that
the constraints one ends up with precisely satisfy the $ISO(1,1)$
algebra.
In fact, in the first order formulation of the
model Strobl considers, one has a non-linear local symmetry
(specific of the model under consideration),
whose generators are responsible for the ``nasty'' constraint algebra.
 As we shall see,
the Poincar\'e gauge theoretical formulation not only
is not the cause of the emerging of the
non-linear algebra, but actually permits
to overcome this problem:
one can introduce an auxiliary condition
that removes the non-linear symmetry without altering
the Poincar\'e gauge symmetry, the diffeomorphism invariance and
the equations of motion.
Therefore,
 our formalism can indeed simplify
the canonical structure of the model  yet preserving all its relevant
symmetries.

Before explaining quantitatively these statements, we would like to
point out a difference between the standard and our approach to gravity
as a gauge theory, difference that was not recognized in Ref. [3].
In Refs. [1, 2], in order to cast gravity as
close as possible to any ordinary non-Abelian gauge theory, we did not
parametrize the translational part of the  Poincar\'e group in such a
way to reproduce general coordinate transformations. Consequently, a
gauge transformation does not entail a coordinate transformation.
Nevertheless, Poincar\'e gauge invariant actions turn out to be
invariant {\it also } under diffeomorphism transformations.

The model we shall deal with is given by the following second-order
Lagrangian [4]
$$ S_{\rm S}= \int d^2x \, {\cal L}_{\rm S} =
\int d^2 x { \sqrt{-g}\over 4}\bigl( \gamma R^2 +
\beta { T}^a{}_{\mu\nu} { T}_a{}^{\mu\nu} +4 \Lambda\bigr)\ \ ,
\eqno(2)$$
where $R$ is the scalar curvature,
$\sqrt{-g}R= \epsilon^{\mu\nu}R_{\mu\nu}=\epsilon^{\mu\nu}(\partial_\mu
\omega_\nu - \partial_\nu \omega_\mu)$
and $\gamma$, $\beta$ and $\Lambda$ are constants.
At this stage, $e^a{}_\mu$ is the {\it zweibein} so that  ${
T}^a{}_{\mu\nu}= \partial_\mu e^a{}_\nu - \partial_\nu e^a{}_\mu
+\epsilon^a{}_b(\omega_\mu e^b{}_\nu -\omega_\nu e^b{}_\mu)$ is the spacetime
torsion.
$S_{\rm S}$ is invariant
under diffeomorphisms and under local Lorentz
transformations.

By introducing the  Lagrange multipliers
$\pi_a$ and $\pi_2$, the action (2) can be rewritten in
a first order form
$$\eqalignno{S_{\rm H}&= \int d^2 x \,  {\cal L}_{\rm H}  =
\int {d^2 x\over 2} \,  \epsilon^{\mu\nu}\left(\pi_2 R_{\mu\nu} +
\pi_a {T}^a{}_{\mu\nu} + \varepsilon_{ab}
e^a{}_\mu  e^b_\nu  E\right)\ \ , & \hbox{(3.a)} \cr
E & =  {1\over 4\gamma} (\pi_2)^2
 -{1\over 2\beta} \pi_a \pi^a -\Lambda
\ \ . & \hbox{(3.b)}\cr}$$
 The action $S_{\rm H}$, contrary to $S_{\rm S}$, presents an
additional non-linear local symmetry
$$\eqalign{\delta_\kappa \pi_2 &= \epsilon_{ab} \pi^a \kappa^b\ \ ,\cr
\delta_\kappa \pi_a &= E\epsilon_{ab} \kappa^b\ \ ,\cr
\delta_\kappa e^a{}_\mu &= - D_\mu \kappa^a -
{1\over \beta} \pi^a \epsilon_{bc} \kappa^b
e^c{}_\mu\ \ ,\cr
\delta_\kappa \omega_\mu &=
{\pi_2\over 2\gamma} \epsilon_{ab} \kappa^a e^b{}_\mu\ \ ,\cr}\eqno(4)$$
where $\kappa^a$ is the infinitesimal local parameter of the symmetry
and $D_\mu \kappa^a = \partial_\mu \kappa^a +\varepsilon^a{}_b \kappa^b
\omega_\mu$.
The symmetry (4) $on-shell$ is  related to the diffeomorphisms [5].
{}From  Eqs. (4) one recognizes that
those terms of the transformation $\delta_\kappa$ that do not depend on
 $(E, \beta, \gamma)$ are just (linear) local $\kappa^a$-translations [2].
Thus, Eqs. (4) are the sum
of  pure translations
plus non-linear transformations that we shall denote by
 $\bar \delta_\kappa$.
The invariance of $S_{\rm H}$ under (4) is
ensured by the fact that
$\delta_\kappa{\cal L}_{\rm H}=
\epsilon^{\mu\nu}\epsilon_{ab}\partial_\mu[(E+2\Lambda)\kappa^a e^b{}_\nu]$.
This symmetry appears in (2) only at the Hamiltonian level, and in this
context it has been extensively  discussed in Ref.[6].

Following  Refs.[1,2],
the action $S_{\rm H}$ can be equivalently rewritten as
 an $ISO(1,1)$ gauge theory
$$\eqalignno{S_{\rm G}&=\int d^2x\,  {\cal L}_{\rm G}  =
\int d^2 x \left[\pi_A F^A{}_{01} + \varepsilon_{ab}
{\cal D}_0
q^a  {\cal D}_1 q^b \tilde E \right]\ \ , & \hbox{(5.a)} \cr
\tilde E & =  {1\over 4\gamma} (\pi_2 -\varepsilon^a{}_b \pi_a q^b )^2
 -{1\over 2\beta} \pi_a \pi^a -\Lambda
\equiv {1\over 4\gamma} (\pi \tilde q)^2
 -{1\over 2\beta} \pi_a \pi^a -\Lambda
\ \ , & \hbox{(5.b)}\cr}$$
 $(\pi_a, \pi_2)\equiv \pi_A $ and
 $ (T^a{}_{\mu \nu} , R_{\mu \nu})\equiv F^A{}_{\mu \nu}$
 being the components of the Lagrange
multiplier and of the field strength along the
Poincar\'e  generators $P_a$ and $J$, respectively
[Notice that in this
 formulation the {\it zweibein} is defined as in (1) and
the spacetime torsion ${\cal T}^a{}_{\mu\nu}$ is given in terms
of the field strength components by ${\cal T}^a{}_{\mu\nu}= T^a{}_{\mu\nu} +
\epsilon^a{}_b q^b R_{\mu\nu}$. For $q^a=0$, ${\cal L}_{\rm G}\equiv
{\cal L}_{\rm H}$ and ${\cal T}^a{}_{\mu\nu}\equiv T^a{}_{\mu\nu}$.]

Besides the diffeomorphism and local Poincar\'e invariances,
 the action  is also
invariant under the following local non-linear symmetry
$$\eqalign{\bar\delta_\kappa \pi_2 &=
 \tilde E q_a\kappa^a\ \ ,\cr
\bar \delta_\kappa \pi_a &= -\tilde E\epsilon_{ab} \kappa^b\ \ ,\cr
\bar\delta_\kappa e^a{}_\mu &= \epsilon_{cd}\kappa^c {\cal D}_\mu
q^d\left({(\pi\tilde q)\over 2\gamma}\epsilon^a{}_b q^b +{1\over
\beta}\pi^a\right)\ \ ,\cr
\bar \delta_\kappa \omega_\mu &= - {(\pi\tilde q)\over 2\gamma}
\epsilon_{ab} \kappa^a {\cal D}_\mu q^b\ \ ,\cr
\bar\delta_\kappa q^a & = \kappa^a\ \ ,\cr}\eqno(6)$$
which is obviously a consequence of the
non-linear symmetry (4) that the model presents in its
first order formulation.  In fact,
up to an overall sign, Eqs. (6) with $q^a=0$ reproduce the non-linear
part of the transformations (4).
Notice that, since $S_{\rm G}$ is {\it also } Poincar\'e gauge invariant,
the translations and the non-linear transformations $\bar\delta_\kappa$
are now independent symmetries.

To simplify  the generator algebra, it is
convenient to consider instead of (5) the equivalent action
$$\eqalignno{S&= \int d^2 x \, {\cal L}=\int d^2 x \, \left[
 p_a \dot q^a + \pi_2 \dot \omega_1 +\pi_a \dot e^a{}_1
+\lambda^a J_a + \omega_0 G_2 + e^a{}_0 G_a\right] &\hbox{(7.a)}\cr
G_2&= \partial_1
\pi_2 + \varepsilon_{ab} \pi^a e^b{}_1 +\epsilon_{ab}p^aq^b\ \ ,
&\hbox{(7.b)}
\cr
G_a&= \partial_1 \pi_a +\varepsilon_{ab} \pi^b \omega_1 + p_a\ \ ,
&\hbox{(7.c)}\cr J_a&= p_a - \tilde E \varepsilon_{ab} {\cal D}_1 q^b\ \ ,
&\hbox{(7.d)}  \cr}$$
where $\lambda^a$ is a Lagrange multiplier transforming as a Lorentz vector
under $ISO(1,1)$ gauge transformations. Eliminating  $p^a$ by
means of the equation of motion, ${ S}$ becomes ${ S}_{\rm G}$.

 From (7) one sees that $\pi_A =(\pi_a, \pi_2)$ and $p_a$ are the momenta
canonically conjugate to $A_1^A = (e^a{}_1 ,  \omega_1)$ and $q^a$,
respectively.
The remaining degrees of freedom  $A_0^A = (e^a{}_0 ,  \omega_0)$
and $\lambda_a$
do not have dynamics:
they play the role of Lagrange multipliers of the ``Gauss' laws''
$G_A=(G_a, G_2)\simeq 0$ and
$J_a\simeq 0$, and
 the definition of their conjugate momenta
 provides 5 primary constraints
[$\pi^{(0)}{}_A\simeq 0$ and $\pi^{(\lambda)}{}_a\simeq 0$,
respectively]. As a consequence,
 the canonical Hamiltonian ${ H}$ will depend
explicitly on the undetermined velocities
$\dot A^A{}_0$ and $\dot \lambda^a$:
 $$ H=\int dx \,  {\cal H}= \int dx\, [ \pi^{(0)}{}_a \dot
e^a{}_0 + \pi^{(0)}{}_2 \dot \omega_0
+\pi^{(\lambda)}{}_a \dot \lambda^a
 -\lambda^a J_a
 - \omega_0 G_2 - e^a{}_0 G_a ]
\ \ .\eqno(8)$$
The ``Gauss' laws'' related to the gauge symmetry are those
associated to the Lagrange multipliers $A^A{}_0$, {\it i.e.} the $G_A$.
In fact, the $G_A$ are the
$ISO(1,1)$ generators  satisfying the Poincar\'e  algebra
$$ \{ G_a (x), G_b (y)\} =0 \quad, \qquad
\{ G_a (x), G_2 (y)\} =\varepsilon_{ab} G^b \delta (x-y)
\ \ .\eqno(9)$$
The remaining  algebra involving  the
  $J_a$ constraints is given by
$$\eqalignno{ \{G_a (x), J_b(y)\} &=0\ \ , &\hbox{(10.a)}\cr
\{J_a (x), G_2(y)\} &=\varepsilon_a{}^b J_b \delta(x-y)\ \ , &\hbox{(10.b)}\cr
\{J_a (x), J_b(y)\} &=
\varepsilon_{ab}\left[{1\over 2\gamma}(\pi\tilde q) (G\tilde q) -{1\over
\beta} \pi^c (G_c -J_c)\right]\delta(x-y)\ \ , &\hbox{(10.c)}\cr}$$
where $(G\tilde q)= G_2-\varepsilon^a{}_bG_aq^b$.
As is apparent from Eqs. (10), the first class algebra of the $J_a$ generators
is non linear, and it contains in the r.h.s. structure functions, rather
than structure constants.
Eq. (10.c) led Strobl to conclude that our Poincar\'e gauge theoretical
formulation of the model is  redundant since ``the constraint algebra is
not just a representation of the Lie algebra of the gauge group''.
This fact, however, should not surprise as  we started from an action that
presents, besides the gauge symmetry, an  additional non-linear local symmetry.
In fact, from Eq. (10.c) it can be easily proved that the constraints
$J_a$ are  precisely the generators of the symmetry (6).
Consequently, they  do not generate diffeomorphisms, as
 alleged in Refs.[3,5]. The transformations (6)  are
related to the diffeomorphisms $on-shell$, but are by no means the same thing.
In fact, as  we shall show,
with the Poincar\'e gauge theoretic formulation
one can break the former symmetry preserving the latter.

The additional
non-linear symmetry can be fixed, so as to eliminate the constraints
$J_a=0$, also
 {\it without}
 breaking the Poincar\'e gauge invariance.

This opportunity is provided by the decoupling of the translations and
of the non-linear symmetry $\bar \delta_\kappa$ that our formalism
entails.
One can in fact  choose
the {\it gauge-covariant} auxiliary conditions
$$ \sigma_a = c\lambda_a -\pi_a \simeq 0\ \ , \eqno(11)$$
where $c$ is any arbitrary constant with dimensions of $length^{-1}$.
Notice that
the constraints (11) can be imposed
due to the introduction of the $q^a$ variables, namely due to the
realization of the model as a gauge theory  (otherwise one would have not
be forced to introduce the $\lambda_a$ variable).
The constraints  (11) make second class the $J_a$ constraints,
so that canonical (Dirac) brackets compatible with the
constraints
$\phi_\alpha = (\sigma_a, J_a)$
strongly equal to zero
 can be consistently defined [7].
 For any pair  ${\cal A}(x), {\cal B}(y)$
of functionals  of canonical variables the Dirac brackets read
$$\eqalign{ \{ {\cal A}(x), {\cal B}(y)\}_{\cal D} & =
\{ {\cal A}(x), {\cal B}(y)\} - \int du \,
\{ {\cal A}(x), \sigma_a(u)\}  {\varepsilon^{ab}\over \tilde E(u)}
 \{ J_b(u), {\cal B}(y)\}\cr & - \int du \,
\{ {\cal A}(x), J_a(u)\}  {\varepsilon^{ab}\over \tilde E(u)}
 \{ \sigma_b(u), {\cal B}(y)\}\ \ .}
\eqno(12)$$
 The constraint algebra in terms of Dirac
brackets then becomes
$$ \{ G_a (x), G_b (y)\}_{\cal D} =0 \quad, \qquad
\{ G_a (x), G_2 (y)\}_{\cal D} =\varepsilon_{ab} G^b  \delta (x-y)
\ \ ,\eqno(13)$$
namely one is left {\it only} with the Poincar\'e algebra.
This is a consequence of the fact that the constraints
 $\sigma_a = 0$ do not violate the gauge symmetry. Moreover the
$\sigma^a=0$,
even if break the $J$-symmetry, do not entail the choice of a coordinate
system, so that they do not break the diffeomorphisms. As a consequence
the equations of motion generated by the Dirac brackets (12)
 with the Hamiltonian $H$
turn out to be
{\it identical} to the equations of motion
obtained from the action given in Eqs. (5),
as can be verified.

In the formulation (3) of the model, to get rid of the
undesirable constraint algebra, one has to break the
symmetry (4).  In this case, however, one looses general covariance and
alters the equations of motion.

Obviously,
by choosing
 the ``physical gauge'' $q^a=0$ instead of the condition (11),
one fixes the translational part of
the Poincar\'e symmetry, and therefore  makes second class the
constraints $G_a$ (or the $J_a$). The remaining generators $J_a$ (or $G_a$) and
$G_2$  and the algebra (13) then  reproduce the generators and the algebra
of the non-linear part of the
 symmetry (4) plus Lorentz ($G_2$)
transformations,
and one  returns to the original model, Eq. (3).
In this case, however,
one is left with an undesirable constraint algebra and the
gauge symmetry is lost.

Thanks to
our method for writing the model as a gauge theory, one can eliminate the
$J_a=0$ constraints maintaining general covariance, and one arrives at
a constraint algebra
that $is$ a representation of the Lie algebra of the gauge group and that
consequently $does$ simplify the Hamiltonian
structure of the model.

The canonical analysis of the Poincar\'e gauge theory  for Liouville
gravity [8] can be performed exactly in the same way (see Ref. [9],
where a more exhaustive discussion
of both the models
is presented).

\medskip
Let us now analyze  the other doubts raised by Strobl. The second main
criticism of Ref.[3] is that our approach -- obtained by introducing the
Poincar\'e coordinates $q^a$ and the Poincar\'e gauge connection $A_\mu$ --
is ``trivially equivalent'', at the classical level,
 to the one obtained by considering the original
action in a first order formalism, with {\it vielbein} $V^a{}_\mu$  and spin
connection as independent variables, and he writes
$$ L(q, e_\mu , \omega_\mu) \sim L (V_\mu , \omega_\mu)\ \ . \eqno(14)$$
We certainly agree on the equivalence, it was our intention to provide
an {\it equivalent} formulation. However,
 the theory defined in the l.h.s. of Eq. (14) is a gauge theory,
the one in the r.h.s. is not.
In particular, in the standard
 first order formalism,
the {\it vielbein} cannot be written in terms of
Poincar\'e gauge potentials $A_\mu$ for which the transformation law is the
standard $\delta A_\mu = -\partial_\mu u - [A_\mu, u]$ [With the exceptions
of pure
 gravity in 3 dimensions [10] and of pure 2-dimensional ``black-hole'' gravity
when treated as a gauge theory of the $extended$ Poincar\'e group [11]].
 To appreciate this point, it is instructive to draw an analogy
with classical electrodynamics. To formulate electrodynamics as a gauge
theory, unphysical extra-degrees of freedom are needed,
the longitudinal components of the gauge potentials.
In some ``physical gauge'' such unphysical components can be
gauged away. In addition, it is well known that
with a suitable shift of gauge potentials,
 the Maxwell action is ``trivially equivalent''
to the action of a massless Proca field. However, in this
case the gauge structure of the Maxwell action  has been unavoidably lost.
 Moreover, in a path integral, the functional determinant of the coordinate
redefinition that leads from the Maxwell to the Proca fields,
in the Lorentz gauge
would  just give an inessential normalization factor, precisely as
the coordinate redefinition proposed by Strobl for our formalism
in the physical gauge.
Nevertheless, for an arbitrary gauge condition $F(q, e_\mu,
\omega_\mu)=0$,
 such redefinition might non-trivially influence the functional
Dirac-$\delta$ of the gauge condition and the Fadeev-Popov determinant,
as it happens in the usual formulation of Yang--Mills theories with an
arbitrary gauge choice $F(A_\mu)=0$ .

The author of Ref.[3], therefore, should not be ``struck'' by the
 redundant and unphysical degrees of freedom that our
formalism entails: this is a characteristic feature of gauge theories.

Other two  remarks in Strobl's comment deserve an answer. The first one
concerns  a technical point related to the possibility of ``neutralizing''
the momentum part of the Poincar\'e transformations in any matter
multiplet $\Phi$ through a suitable redefinition of $\Phi$. We explained
exhaustively this point at the end of Sect. 3 in Ref.[2].

Strobl considers, as an example, the Poincar\'e gauge invariant scalar field
action in 4 dimensions that we provided in Ref.[1].
He uses what we called in Ref.[2] the transformation to the zero momentum
representation.

 Following Ref.[2], we can be
more general by considering any matter  multiplet in any dimension:
if $\Phi$ is a matter multiplet transforming according to
any given representation (J,P) of the Poincar\'e group,  then the
multiplet $\tilde \Phi = ({\bf 1} - q^a P_a)\Phi$ transforms according
 to the zero-momentum representation (J,0), namely
$$ \delta \Phi^A = (\alpha \cdot J + \rho \cdot P)^A{}_B \Phi^B
\Longrightarrow \delta \tilde\Phi^A = (\alpha \cdot J)^A{}_B \tilde\Phi^B
\ \ .\eqno(15)$$
{}From this property Strobl concludes that the translational part of the
Poincar\'e gauge group is superfluous in the gauge theory, and
that the theory is equivalent to the one obtained
by considering the matter multiplet in the zero-momentum representation
and the {\it vielbein} $V^a{}_\mu$ as an independent variable.
The theory one gets is certainly equivalent at the classical level,
but it is no  longer a gauge theory, because in this case the {\it vielbein}
is not expressed in terms of Poincar\'e gauge potentials.

The last of Strobl's comments concerns an incorrect statement in
the Appendix of Ref.[2]. He has an old version
of our paper.
In fact, in the first release of the preprint  the Appendix was mistaken,
but it was promptly replaced by a correct one about
a week after the submission: at least since October 1992,
in the version of our preprint
 that can be found in all the electronic libraries,
 the Appendix is correct.

Finally, we take the opportunity to make a remark on our papers. In
Refs.[1,2,9,12] the whole  analysis was performed at the classical level.
Concerning possible developments at the quantum level, we always used the
conditional form. Certainly we do not expect that our procedure solves
the huge problems that arise in quantum gravity.
The real problems  of a quantum theory of gravity presumably
cannot be overcome  just  by  a gauge theoretical description of the theory.
Nevertheless, such a description could provide
 an alternative to the standard approach that could be worth
to investigate.

\bigskip
\bigskip

\noindent{\bf REFERENCES}
\medskip
\nobreak
\bigskip

\item{[1]} G. Grignani and G. Nardelli, {\it Phys. Rev. } {\bf D45}
(1992) 2719.
 \medskip
\item{[2]}  G. Grignani and G. Nardelli, ``Poincar\'e Gauge Theories for
Lineal Gravities'',  Preprint
DFUPG-57-1992/UTF-266-1992.
\medskip
\item{[3]} T. Strobl, ``Comment on Gravity and the Poincar\'e Group'',
Preprint TUW-01-1993.
\medskip
\item{[4]} M. O. Katanaev and I. V. Volovich, {\it Ann. Phys.} (N.Y.)
 {\bf 197} (1986) 1.
\medskip
\item{[5]} T. Strobl, {\it Int. J. Mod. Phys.} {\bf 8}, (1993) 1883.
\item{[6]}H. Grosse, W. Kummer, P. Presnaider and D. J. Schwarz,
{\it J. Math. Phys.} {\bf 33} (1992) 3892.
\medskip
\item{[7]} See for example A. Hanson, T. Regge and C. Teitelboim,
``Constrained Hamiltonian Systems'', Accademia Nazionale dei Lincei,
Roma 1979.
\medskip
\item{[8]}C. Teitelboim, {\it Phys. Lett. } {\bf B126}, 41 (1983), and
in {\it Quantum Theory of Gravity}, S Christensen ed. (Adam Higler,
Bristol, 1984); R. Jackiw in
{\it Quantum Theory of Gravity}, S Christensen ed. (Adam Higler,
Bristol, 1984) and {\it Nucl. Phys. } {\bf B252}, 343 (1985).
For a first order formulation of Liouville gravity as a de--Sitter
gauge theory see for instance K. Isler and C. A. Trugenberger, {\it
Phys. Rev. Lett.} {\bf 63}, 834 (1989).
\medskip
\item{[9]}G. Grignani and G. Nardelli, ``Canonical Analysis of
Poincar\'e Gauge Theories for Two Dimensional Gravity'', Preprint
DFUPG-74-1993/UTF-292-1993, March 1993.
\medskip
\item{[10]}A. Achucarro and P. Townsend, {\it Phys. Lett.} {\bf B180}
(1986) 85; E. Witten, {\it Nucl. Phys.} {\bf B311} (1988) 46.
To this purpose it should be noticed that even in 3 dimensions, if
matter interactions are present, the component of the gauge potential
along the momentum generators cannot be interpreted as {\it dreibein}
without encountering inconsistencies.
\medskip
\item{[11]}D. Cangemi and R. Jackiw, ``Poincar\'e Gauge Theory for
Gravitational Forces in 1+1 Dimensions'', MIT preprint CTP \# 2165 ({\it
Ann. of Phys.}, in press) and references therein.
\item{[12]}G. Grignani and G. Nardelli, {\it Phys. Lett.} {\bf B264}
(1991) 45; ibid, {\it Nucl. Phys.} {\bf B370} (1992) 491; ibid,
{\it Phys. Lett.} {\bf B300} (1993) 38.
\vfill
\eject

\end